\documentclass[aps,prx,twocolumn,longbibliography,groupedaddress,floatfix,superscriptaddress]{revtex4-1}
\usepackage{graphicx}  
\usepackage{dcolumn}   
\usepackage[english]{babel}
\usepackage{bm}        
\usepackage{amsmath, amsthm, amssymb}
\usepackage{bbold} 
\usepackage{mathrsfs}
\usepackage{array}
\usepackage[usenames]{color}
\usepackage{soul} 
\usepackage{ulem}
\usepackage{comment}

\emergencystretch=\maxdimen
\begin{document}

\title{Interplay of Flat Electronic Bands with Holstein Phonons}
\author{Chunhan Feng} 
\affiliation{Department of Physics, University of California,
Davis, CA 95616,USA}
\author{Richard T. Scalettar}
\affiliation{Department of Physics, University of California, 
Davis, CA 95616,USA}

\begin{abstract}
Existing Quantum Monte Carlo studies have investigated the
properties of fermions on a Lieb (CuO$_2$) lattice interacting with an
on-site, or near-neighbor electron-electron coupling.
Attention has focused on the interplay of
such interactions with the macroscopic degeneracy of local
zero energy modes, from which Bloch states can be formed
to produce a flat band in which energy is independent of
momentum.  The resulting high density of states, in combination
with the Stoner criterion, suggests that there should be pronounced
instabilities to ordered phases.  Indeed, a theorem by Lieb rigorously
establishes the existence of ferrimagnetic order.  Here we study the
charge density wave phases induced by electron-phonon coupling
on the Lieb lattice, as opposed to previous work
on electron-electron interactions.
Our key result is the demonstration of charge density wave 
(CDW) phases at
one-third and two-thirds fillings, characterized by long-range density density
correlations between doubly occupied sites on the minority or majority sublattice,
and an accompanying gap.  We also compute the transition temperature to the ordered
phase as a function of the electron-phonon coupling.
\end{abstract}

\date{\today}

\maketitle

%%%%%%%%%%%%%%%%%%%%%%%%%%%%%%%%%%%%%%%%%%%%%%%%%%%%%%%%%%%%%%%%%%
\section{Introduction}  \label{sec:Introduction}
%%%%%%%%%%%%%%%%%%%%%%%%%%%%%%%%%%%%%%%%%%%%%%%%%%%%%%%%%%%%%%%%%%

A number of periodic tight-binding lattices contain 
a macroscopic degeneracy of local, zero energy eigenstates which 
arise from the perfect cancellation of hopping for an appropriately
phased occupation state\cite{derzhko15,leykam18}.
These include the Kagom\'e, sawtooth, Creutz,
diamond-octagon, square-octagon, decorated honeycomb, and finally the
dice lattice, where the phenomenon was first noted\cite{sutherland86}.
One of the most prominent examples is the Lieb lattice,
shown in Fig.~\ref{fig:lattice_structure}, which is of
special interest as the structure of the CuO$_2$ planes of the
cuprate superconductors.

The existence of these `compact localized states' is a property
of the non-interacting system.  Several years after their
discovery, it was pointed out that precise statements 
can be made concerning the role of repulsive electron-electron
interactions in flat band systems.
Specifically, the existence of a ferrimagnetic ground state
can be rigorously established\cite{Lieb89}.
Subsequent work further investigated flat band ferromagnetism
\cite{mielke91,Mielke1991b,tasaki92,tasaki98}.
The effect of attractive electron-electron interactions is
also of interest\cite{julku16,kumar19,huhtinen20,swain20}, especially since the momentum at which
Bose-Einstein condensation of fermionic pairs might occur is uncertain
in a flat band\cite{huber10,tovmasyan13,iglovikov14}.

Flat bands have also been considered within the context of
the Fractional Quantum Hall Effect,
\cite{parameswaran13}
Chern insulating behavior,
\cite{bergholtz13},
Tomonaga-Luttinger liquids
\cite{takayoshi13}
and Haldane phases 
\cite{gremaud17}.  Perhaps the most dramatic explosion of 
theoretical and computational interest
coincided
with the recent discovery that bi-layer graphene, when twisted at a 
``magic angle" of about 1.1 degrees, displays unconventional 
superconductivity (SC) which is likely closely linked to the appearance of a 
nearly dispersionless bands in the effective Moire pattern lattice
\cite{cao18,cao18b,guo18,pinto20,lee20,shen20}.
This SC is characterized by a ratio of
critical temperature to Fermi temperature higher than the cuprates.

In addition to realizations in these solid state
materials, flat band physics has also been explored in 
photonic Lieb Lattices\cite{GuzmanSilva14,Mukherjee15},
and optical Lieb \cite{Noda14,Xia16},
Kagom\'e\cite{santos04}
and honeycomb\cite{wu07} lattices.

Here, we investigate the phases of interacting 
{\it electron-phonon} systems 
for flat electronic bands\cite{li20}.  Specifically, we study the Holstein Hamiltonian
on a Lieb lattice.  
Although there are suggestive analogies between the Holstein model
and the attractive Hubbard model, the former has a non-trivial frequency
dependent coupling which distinguishes the two situations, the most significant
consequence of which is the presence of a finite temperature 
phase transition even on 2D lattices which are the most commonly
investigated flat band geometries.
It is only in the extreme anti-adiabatic limit, where the
phonon frequency is one to two orders of magnitude larger than
the electronic bandwidth, that the Holstein and attractive Hubbard models
become quantitatively equivalent\cite{feng20}.

The structure of this paper is as follows:
After introducing the model 
(Sec.~\ref{sec:HolsteinModel}) and computational
methodologies (Sec.~\ref{sec:MFTandDQMC}),
we show the behavior of the compressibility, double occupancy, spectral
function, and charge density wave structure factor
(Sec.~\ref{sec:Results}).  Together these observables point to
the formation of a gapped charge density wave (CDW) state below a critical temperature $T_c$,
whose value we determine using finite size scaling.
A final section summarizes our findings.

%%%%%%%%%%%%%%%%%%%%%%%%%%%%%%%%%%%%%%%%%%%%%%%%%%%%%%%%%%%%%%%%%%
\section{Holstein Model}  \label{sec:HolsteinModel}
%%%%%%%%%%%%%%%%%%%%%%%%%%%%%%%%%%%%%%%%%%%%%%%%%%%%%%%%%%%%%%%%%%

\begin{figure}[t]
\includegraphics[height=2.7in,width=3.5in]{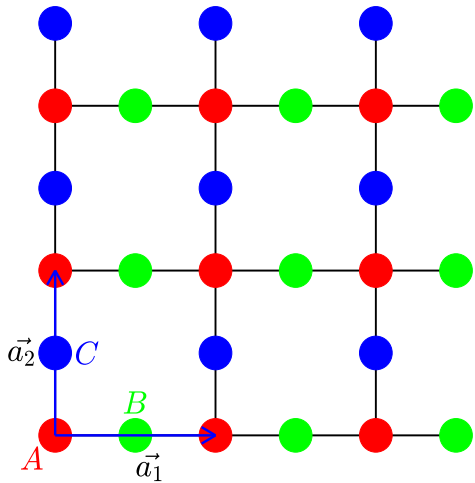}
\caption{The Lieb lattice geometry.  Additional sites (blue and green) 
are added to midpoint of each of the bonds linking the 
sites of a square lattice (red).  The resulting structure is bipartite and
has three sites per unit cell.  Note especially that the red sublattice
contains only half as many sites as the sublattice comprised of blue
and green sites. The blue/green pattern of sites
surrounding one of the vacancies illustrates a zero energy
mode. See text.
}
\label{fig:lattice_structure} 
\end{figure}

The Holstein model\cite{holstein59}
we consider consists of a collection of electrons, described
by fermionic creation and destruction operators 
$\hat d^{\dagger}_{\mathbf{i} \sigma},
\hat d^{\phantom{\dagger}}_{\mathbf{i} \sigma}$
hopping between near neighbor sites on the Lieb lattice
shown in Fig.~\ref{fig:lattice_structure}.
The electron density on each site,
$\hat n_{\mathbf{i}} = 
\hat n_{\mathbf{i}\uparrow} + 
\hat n_{\mathbf{i}\downarrow}$ with
$
\hat n_{\mathbf{i}\sigma} =
\hat d^{\dagger}_{\mathbf{i} \sigma} 
\hat d^{\phantom{\dagger}}_{\mathbf{i} \sigma}$, where $i$ denotes lattice sites and $\sigma$ is the spin index,
couples linearly to the displacement $\hat x_{\mathbf{i}}$
of a local quantum oscillator degree of freedom.
The Hamiltonian is therefore,
\begin{align} 
\nonumber \mathcal{H} = & -t \sum_{\langle \mathbf{i}, \mathbf{j}
  \rangle, \sigma} \big(\hat d^{\dagger}_{\mathbf{i} \sigma}
\hat d^{\phantom{\dagger}}_{\mathbf{j} \sigma} + {\rm h.c.} \big) -
\mu^{\phantom{\dagger}} 
\sum_{\mathbf{i}, \sigma} 
\hat n^{\phantom{\dagger}}_{\mathbf{i} \sigma} 
\\& +\frac{1}{2} \sum_{\mathbf{i}} \big( \, \hat p_{\mathbf{i}}^2
 + \omega_0^2 \hat x_{\mathbf{i}}^2 \,\big)
 + \lambda \sum_{\mathbf{i},\sigma} \hat x_{\mathbf{i}} 
 \hat n_{\mathbf{i}\sigma}
~~ .
\label{eq:ham}
\end{align}
We have set the oscillator mass $M=1$ and will also use units in which 
$\hbar=k_{\rm B}=1$ and the hopping amplitude $t=1$.
The chemical $\mu=-\lambda^2/\omega_0^2$ corresponds to half filling. 

The Lieb lattice Hamiltonian is sometimes studied with an additional  `charge transfer' 
term in the form of an energy difference between the sites on the
two sublattices.  We do not include such a term here.
Its inclusion would favor one of the two degenerate CDW phases and preempt
the spontaneous symmetry breaking phase transition which is our focus here.

\begin{figure}[t]
\includegraphics[height=2.35in,width=3.2in]{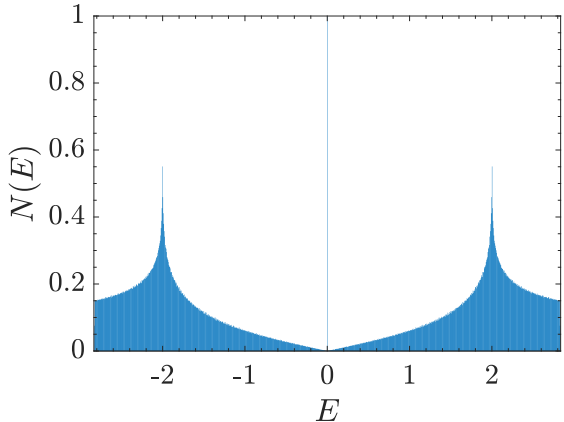}
\caption{The density of states of the Lieb lattice.  Energy levels
of two dispersing bands bracket the $\delta$-function peak at $E=0$.
Particle-hole symmetry is reflected in the property that
$N(E)=N(-E)$.
}
\label{fig:DOS} 
\end{figure}

\begin{figure}[t]
\includegraphics[height=2.35in,width=3.2in]{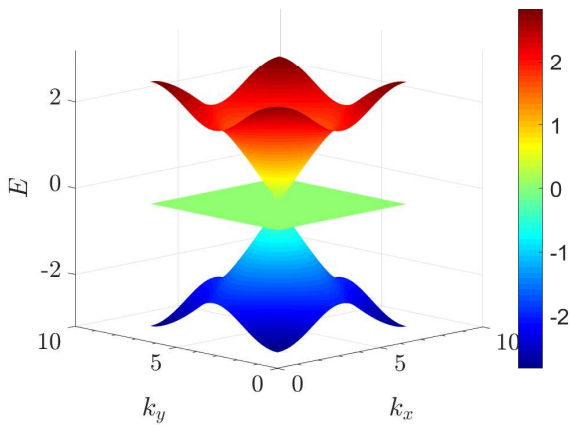}
\caption{The band structure of the Lieb lattice.}
\label{fig:spectrum} 
\end{figure}

The electronic density of states 
in the absence of the electron-phonon interactions, 
is given in Fig.~\ref{fig:DOS}.
The $\delta-$function spike at $E=0$ reflects the macroscopic degenerate
collection of local $E=0$ 
vectors $|\psi\rangle$ constructed by forming
a state with equal amplitude and
opposite phases on the four blue and green sites surrounding any vacant site on the
Lieb lattice.  See Fig.~\ref{fig:lattice_structure}.
All these $|\psi\rangle$ 
have the property $\hat {\cal K} |\psi \rangle=0$.
where $\hat {\cal K}$ is the first (hopping) term in Eq.~\ref{eq:ham}.
The band structure is given in Fig.~\ref{fig:spectrum} 

When $\lambda \neq 0$,
the qualitative physics of the Holstein model is as follows:
at low densities individual electrons distort the lattice sites in
their vicinity.  The resulting composite particle, a `polaron',
possesses an increased effective mass, reflecting the fact that when
the electron hops between sites, 
the oscillator degrees of freedom must reconfigure 
themselves\cite{Freericks93,romero99,ku02,hohenadler04,marchand13}.
These dressed quasiparticles tend to attract one another, since
the distortion caused by one provides a favorable environment
for another.  Indeed, solving the $t=0$ Holstein model
one can see an effective attraction
$U_{\rm eff} = -\lambda^2/\omega_0^2$ exists between spin up
and down fermions.  This independent site form is consistent with 
the interaction between electrons mediated by a phonon
propagator, $V_{\rm eff}(\omega) = \lambda^2/
(\, \omega^2 - \omega_0^2 \,)$, if one sets $\omega=0$.

The pairs of up and down electrons which arise from
this attraction can participate in ordered phases.  One possibility,
which dominates on half-filled 
($n_{\mathbf{i},\sigma}=1/2$)
bipartite lattices with equal number of
sites in the two sublattices,
such as square and honeycomb geometries,
is a CDW arrangement in
which pairs occupy one of the two sublattices.  
CDW formation is energetically favorable because, by surrounding
itself with empty sites, a pair of electrons has
the optimal ability for virtual hopping processes to
adjacent sites, thereby lowering its energy by
$J \sim - z t^2/U_{\rm eff}$ where $z$ is the coordination number.
This situation is similar to that giving rise to antiferromagnetic
order in the half-filled repulsive Hubbard model.   

Another possible ordered state occurs when the pairs
condense into a superconducting phase.  This is expected to occur
when the system is doped away from fillings which favor CDW order
and has been studied with, for example, Eliashberg theory
\cite{scalettar89b,marsiglio92,vonderlinden95,alexandrov01,chubukov19,dee2020relative}.
QMC simulations have given indications of pairing 
as well\cite{noack91,vekic92,bradley20}.

In this paper, we consider the CDW transition in the Holstein
model on the Lieb lattice.  We set the phonon frequency $\omega_0/t=1$
to facilitate comparisons with most of the existing 
QMC literature\cite{Scalettar89,noack91,vekic92,vekic93,weber18,costa18,zhang19,cohenstead19,chen19,xiao20}.
This historical choice was in part made as a simple starting point
to explore the qualitative physics of the CDW and SC transitions,
but also because it facilitated the Determinant
Quantum Monte Carlo (DQMC) simulations, which were known to
exhibit long autocorrelation times at $\omega_0/t \lesssim 1/2$.
Recent algorithmic improvements have made possible the study of smaller
$\omega_0$\cite{chuang18,beyl2018,batrouni19a,batrouni19b,zhang20}.

%%%%%%%%%%%%%%%%%%%%%%%%%%%%%%%%%%%%%%%%%%%%%%%%%%%%%%%%%%%%%%%%%%
\section{Computational Methodologies}  \label{sec:MFTandDQMC}
%%%%%%%%%%%%%%%%%%%%%%%%%%%%%%%%%%%%%%%%%%%%%%%%%%%%%%%%%%%%%%%%%%

\subsection{Mean Field Theory}

We use the adiabatic approximation, ignoring the
$\hat p_{\bf i}^2$ term, and assume a staggered pattern of phonon 
displacements with the ansatz $x_{\bf i} = x_0 - \Delta$ 
for sublattice A and $x_{\bf i} = x_0 + \Delta$ for sublattice 
B/C. Inserting this ansatz into Eq.~\ref{eq:ham}, the resulting
quadratic fermion Hamiltonian can be diagonalized. Then the 
free energy is a function of $x_0$, $\Delta$ and inverse temperature $\beta$,
\begin{align}
F =\frac{1}{2} N \omega_0^2 (x_0^2+\Delta^2 + \frac{2}{3} x_0
\Delta) - \frac{1}{\beta} \sum_{\alpha,\sigma,\mathbf{k}}
{\rm ln} \, (1+e^{-\beta \epsilon_{\alpha}}),
\label{eq:free energy}
\end{align}
where 
\begin{align*}
\begin{split}
\epsilon_{\alpha}= \left \{
\begin{array}{ll}
\lambda \Delta+\lambda x_0 -\mu, &\\
 \pm 
\sqrt{(\lambda \Delta)^2+4t^2({\rm cos^2}\frac{k_x}{2}+{\rm cos^2}\frac{k_y}{2})}+\lambda x_0 -\mu     &\\
\end{array}
\right.
\end{split}
\end{align*}
 are the three fermion energy bands, and $\mathbf{k}=(k_x,k_y)$ 
are allowed momentum vectors. At a fixed temperature $T$, we 
determine the ($x_0^*$, $\Delta^*$) which minimize $F$. 
Results obtained by this approach will be presented in the next section.

\subsection{Determinant Quantum Monte Carlo}

Although much insight can be gleaned from MFT, especially concerning the
possible types of order, it has a number of well-understood defects,
especially an overestimate of the tendency to long range order arising
from ignoring fluctuations.  This is particularly evident in
lattice models like the Hubbard and Holstein Hamiltonians
where it fails to distinguish two separate energy
scales.  The first is the temperature $T\sim U$ at which local 
moments (in the case of repulsive interactions) or pairs 
(in the case of attractive interactions) form.  
The second is the temperature at which inter-site ordering
occurs.  Since the former grows linearly with the interaction
strength $U$, and the latter falls as $1/U$, MFT 
overestimates $T_c$ by a far wider margin at strong coupling
than in simpler classical descriptions of long range order
such as the Ising model.

To provide a more accurate treatment of the electron-phonon
correlations, we turn to the use of the DQMC methodology\cite{blankenbecler81,sorella89}.
In this approach, the full imaginary time propagator 
$e^{-\beta \hat {\cal H}}$ is written as a product of incremental
factors $e^{-\Delta \tau \hat {\cal H}}$.
This discretization allows for the `Trotter' approximation,
$e^{-\Delta \tau \hat {\cal H}} \approx
e^{-\Delta \tau \hat {\cal H}_1}
e^{-\Delta \tau \hat {\cal H}_2}$
with 
$\hat {\cal H} = \hat {\cal H}_1 + \hat {\cal H}_2$.
The purpose of dividing up the imaginary time evolution is that
the matrix elements of the individual pieces can be evaluated analytically.
In particular, upon the introduction of complete sets of phonon states, the
fermionic trace in the  resulting quadratic form of fermionic
operators can be performed, leaving
a trace over a phonon field $x(\mathbf{i},\tau)$ which depends on
both spatial site $\mathbf{i}$ and imaginary time slice $\tau$.
The integrand has both a bosonic piece from the quantum oscillator
term in $\hat {\cal H}$ and a product of two determinants
(one from each spin species) which depend on 
$x(\mathbf{i},\tau)$.
For the Holstein model, because the up and down species couple to the phonon coordinate in the same way, the determinants are  identical. 
The
fermion sign problem is absent in the resulting square of determinants.
$x(\mathbf{i},\tau)$ is sampled stochastically.

DQMC treats interacting quantum Hamiltonians exactly.  The sole
(controlled) approximation is in the discretization of $\beta$.  With the 
usual choices of $\Delta\tau$ the associated errors are easily made smaller
than those arising from the sampling.  (The exception is for local
quantities like the energy and double occupancy whose statistical
errors are extremely small.  For these observables, a
$\Delta \tau \rightarrow 0$ extrapolation is straightforward to perform.)
Simulations are carried out on lattices of finite size, necessitating
a finite size scaling analysis, as described below.

We focus on several local observables, the density 
$\rho = \langle \hat n_{\mathbf{i}} \rangle$ and double occupancy
${\cal D} = \langle \hat n_{\mathbf{i}\uparrow} 
\hat n_{\mathbf{i}\downarrow} \rangle$, and on the CDW structure factor, the
Fourier transform of the real-space density-density correlation function.
\begin{align}
S(\mathbf{q}) &= \sum_{\mathbf{r}} c(\mathbf{r}) \, e^{i \mathbf{q} \cdot \mathbf{r}}
\nonumber \\
c(\mathbf{r}) &= \langle \, \Delta \hat n_{\mathbf{i+r}} 
\, \Delta \hat n_{\mathbf{i}}\, \rangle
\,\, ,
\label{eq:Sq}
\end{align}
where $\Delta \hat n_{\mathbf{i}}=\sum_{\sigma} \Delta \hat n_{\mathbf{i},\sigma}=\sum_{\sigma} \hat n_{\mathbf{i}B,\sigma}+
\hat n_{\mathbf{i}C,\sigma}-2\hat n_{\mathbf{i}A,\sigma}$ is the 
charge density difference within a unit cell, labeled by
${\bf i}$.
\textcolor{black}{
When only the A or B/C sublattice is occupied,
corresponding to one-third or two-thirds filling,
the dominant $S(\mathbf{q})$ will be $S_{\rm cdw} = S(0,0)$.
}

The spectral function $A(r,\omega)$ is obtained by an analytic continuation of the
non-equal time Greens function
\begin{align}
G(\mathbf{r},\tau) &=  \langle \, \hat c_{\mathbf{i+r},\sigma}(\tau) 
\hat c_{\mathbf{i},\sigma}(0) \, \rangle
\nonumber \\
&=   \langle \, e^{\tau \hat {\cal H}} \hat c_{\mathbf{i+r},\sigma}(0) e^{-\tau \hat {\cal H}} 
\hat c_{\mathbf{i},\sigma}(0) \, \rangle
\nonumber \\
G(\mathbf{r},\tau) &=  \int d\omega  A(\mathbf{r},\omega) 
\frac{e^{-\omega \tau}}{e^{\beta \omega}+1}
\label{eq:Aw}
\end{align}
We report the Fourier transform of the spectral function at zero distance,
a quantity which is the analog of the non-interacting density of states
in a correlated system.

DQMC has been used to explore various properties of the attractive and 
repulsive Hubbard
models on the Lieb Lattice\cite{iglovikov14,Costa16,Oliveira19},
but has not yet been used for the Holstein model.

%%%%%%%%%%%%%%%%%%%%%%%%%%%%%%%%%%%%%%%%%%%%%%%%%%%%%%%%%%%%%%%%%%
\section{Results}  \label{sec:Results}
%%%%%%%%%%%%%%%%%%%%%%%%%%%%%%%%%%%%%%%%%%%%%%%%%%%%%%%%%%%%%%%%%%

\subsection{Mean Field Theory}

\begin{figure}[t!]
\includegraphics[height=2.35in,width=3.2in]{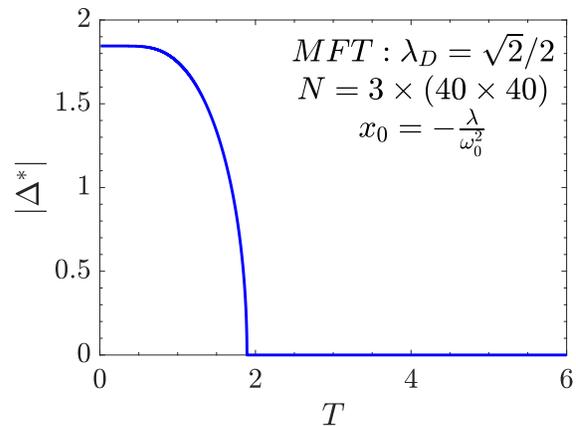}
\caption{
Mean field order parameter
$\Delta^*$ as a function of temperature $T/t$
at half-filling, $\mu=-\frac{\lambda^2}{\omega_0^2}=-4$. 
Here and in all subsequent figures $\omega_0/t=1$.
For $T > T_c \sim 1.9\,t$, the MFT critical temperature,
$\Delta^*=0$ and each site has $\rho_{\bf i}=1/2$ per spin.
For $T<T_c$ there are two 
degenerate values of $\Delta=\pm \Delta^*$ which minimize
${\cal F}$. 
These correspond to $1/2-d\rho$ and
$1/2+d\rho$ (and hence the {\it average} density is
half-filled).  (See Fig.~\ref{fig:MFT_n_vs_T} and text for more discussion).
 }
\label{fig:MFT_Delta_vs_T} 
\end{figure}

\begin{figure}[t!]
\includegraphics[height=2.35in,width=3.2in]{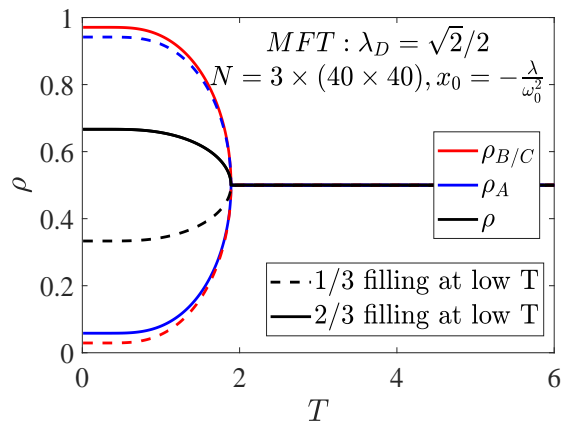}
\caption{
Black full ($-\Delta^*$) and dashed ($+\Delta^*$) curves denote the
electron density per spin on the whole lattice, $\rho$. 
Blue and red colors give densities on the two sublattices $\rho_A$ and $\rho_{B/C}$.
The horizontal axis is temperature $T$.
For $T > T_c \sim 1.9\,t$, the MFT critical temperature, each sublattice has $\rho_{A}
=\rho_{B/C}=1/2$ per spin.
For $T<T_c$, there are two degenerate states.  The densities bifurcate into two
curves associated with the pair of degenerate values $\pm \Delta^*$ of the
order parameter.
 }
\label{fig:MFT_n_vs_T} 
\end{figure}

\begin{figure}[t!]
\includegraphics[height=2.35in,width=3.2in]{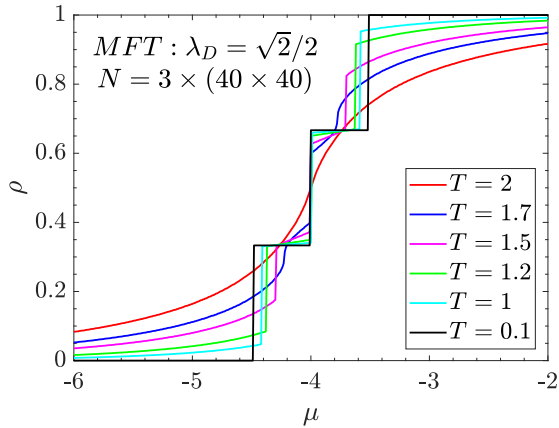}
\caption{
Density per spin $\rho$ as a function of chemical potential $\mu$ within
MFT.  
Temperature $T=2\,t > T_c$ and $\rho(\mu)$ is smooth.  
For temperature $T=\,t < T_c$ the density $\rho(\mu)$ (per spin) has plateaus
at $\rho=1/3, 2/3$ corresponding the a non-zero CDW gap.  %% }
 The sublattice spin occupations are shown in Fig.~\ref{fig:MFT_n_vs_T}. 
When $\Delta^*>0$ there is a smaller number of
$A$ sites with $\rho>1/2$ and a larger number of $B/C$ sites
with $\rho<1/2$ and the total density $\rho \sim 1/3$ (see text), and vice-versa
for $\Delta^*<0$.
}
\label{fig:MFT_n_vs_miu} 
\end{figure}

\begin{figure}[t!]
\includegraphics[height=2.35in,width=3.2in]{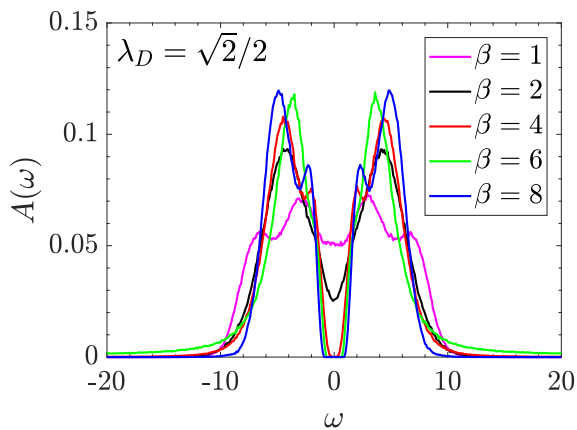}
\caption{
The spectral function $A(\omega)=\frac{1}{N} \sum_k A(k,\omega)$ determined in DQMC calculations.
A gap opens at the Fermi surface $\omega=0$ 
as the temperature is lowered ($\beta$ increases).
This provides a rough estimate of $T_c$.
}
\label{fig:Aw} 
\end{figure}
We first explore the effect of electron phonon interaction by 
using the mean field theory approach described in section 3A. 
Since the $\lambda x_{\bf i} n_{\bf i}$ term in the mean 
field Holstein Hamiltonian can be viewed as a chemical 
potential  $\lambda x_{\bf i}$  acting on site {\bf i}, 
a nonzero bond dimerization $\Delta$ implies a staggered pattern 
of electron density, i.e.~a CDW phase. We set
$\mu=-\frac{\lambda^2}{\omega_0^2}$ so that the lattice is 
half filled.  The corresponding $x_0=-\frac{\lambda}{\omega_0^2}$.

The value $\Delta^*$ which minimizes ${\cal F} $
is plotted as a function of temperature $T$ in 
Fig.~\ref{fig:MFT_Delta_vs_T}. 
For $T>T_c \sim 1.9\,t$, the order parameter $\Delta^*=0$
and there are equal sublattice densities $\rho_{A}=\rho_{B/C}=1/2$ per spin.
(See also Fig.~\ref{fig:MFT_n_vs_T}.)
Below $T_c$, we find there is a degenerate pair of nonzero solutions
$\pm \Delta^*$, and distinct 
densities $\rho_{A};\,\rho_{B/C}$ on the two sublattices. 
We denote the densities per spin on the {\it whole} lattice,
i.e.~averaged over sublattices, 
$(\rho_{A} + 2 \rho_{B/C})\,/\,3$,
by $1/2 \pm d\rho$.  The two signs are associated with the
two signs $\pm \Delta^*$.
A change in sign of $\Delta^*$ can be viewed an interchange 
$A \leftrightarrow B/C$ of the high and low occupation sublattices.
Since the numbers of sites in the two sublattices are unequal,
this also shifts the density on the whole lattice (unlike the more conventional
cases of square and honeycomb bipartite lattices).

Perfect CDW order, in which 
$1/2 \pm d\rho =1/3; \, 2/3$, and 
$(\rho_A; \, \rho_{B/C}) = (1,0)$ or $(0,1)$,
requires the absence of both thermal ($T \rightarrow 0$) and quantum ($\lambda^2/\omega_0^2 \rightarrow \infty$) fluctuations.
In Fig.~\ref{fig:MFT_Delta_vs_T}, 
 $\Delta^*$  increases to a maximal value
 $\Delta^* \sim 1.85$ at zero temperature.  
 $\Delta^*=2$ would yield a perfect CDW pattern.
 That $\Delta^*<2$ reflects the
 presence of some residual quantum fluctuations: $\lambda^2/\omega_0^2$ is finite.
Not surprisingly,  in Fig.~\ref{fig:MFT_n_vs_T},
the density per spin $\rho_{B/C}$ (red) is closer to the perfect CDW state, 
$\rho=0$ (empty) or $\rho = 1$ (doubly occupied), than the density $\rho_{A}$ (blue). 
This is because sites A have twice as many nearest neighbors as sites B/C.  The larger
number of hoppings $t$ produce more quantum fluctuations.
 
\color{black}
All MFT results presented 
in this paper are obtained on a $3\times(40\times40)$ 
Lieb lattice with a dimensionless electron phonon coupling  constant 
$\lambda_D \equiv \frac{\lambda^2}{\omega_0^2 W}=\sqrt{2}/2$. 
Here $W=4 \sqrt{2}t$ is the fermion band width for a Lieb
lattice in the noninteracting limit. We will see later the MFT
$T_c \sim 1.9\,t$ is more than
an order of magnitude higher than the $T_c$ given by DQMC.

For different chemical potential $\mu$, we follow the 
same steps to determine $(x_0^*,\Delta^*)$ minimizing the 
free energy and find $\Delta^*>0$ ($\rho=1/3$ CDW pattern) 
when $\mu<-\frac{\lambda^2}{\omega_0^2}$; $\Delta^*<0$ 
($\rho=2/3$ CDW pattern) when $\mu>-\frac{\lambda^2}{\omega_0^2}$.
The electron density can be obtained by
$n=\sum_{\alpha,\mathbf{k}} 
\frac{1}{1+e^{\beta \epsilon_{\alpha}}}$. 
Figure \ref{fig:MFT_n_vs_miu} shows the density $\rho$ per spin 
as a function of chemical potential $\mu$. As temperature 
is lowered, plateaus at $\rho=1/3$ and $\rho=2/3$ develop, 
indicating that a 1/3 filling CDW pattern and its partner 
at 2/3 filling, 
extend over a finite range of $\mu$, which is consistent 
with the DQMC results below. A similar phenomenon is also 
observed in the `$t-V$ model' of spinless fermions interacting
with a nearest 
neighbor repulsion on a Lieb lattice.\cite{PhysRevB.95.035108}.

\color{black}

\subsection{Determinant Quantum Monte Carlo}

We now turn to DQMC results.  We begin with the spectral function in Fig.~\ref{fig:Aw}.
At high temperatures (small $\beta$) $A(\omega=0)$ is non-zero.  
A gap is fully formed at $\beta_c\,t \sim 6$, suggesting a transition to an insulating CDW phase.

\begin{figure}[t!]
\centering
\includegraphics[height=2.35in,width=3.6in]{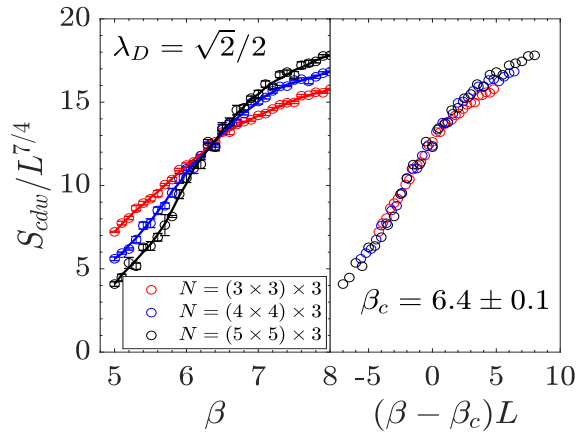}
\caption{
\underbar{Left:} The scaled structure factor is 
plotted versus $\beta$ for three lattice
sizes.  The crossing gives the position of the CDW transition.  
\underbar{Right:}  If the horizontal (inverse temperature) 
axis is also scaled, a full data collapse is obtained.
}
\label{fig:fss} 
\end{figure}

A more accurate determination of the location of the  CDW 
transition is obtained by a finite size scaling analysis of 
$S_{\rm cdw}$.  Because the low temperature phase involves 
occupying one of two spatial sublattices,
it breaks a ${\cal Z}_2$ symmetry, and therefore the 
transition should be in the Ising universality class.  
Using the known 2D Ising critical exponents $\nu=1$
and $\gamma/\nu=7/4$ yields the finite size scaling 
plots of Fig.\ref{fig:fss}.
We find $\beta_c \,t=6.4 \pm 0.1$.
If we eschew this knowledge and instead vary the critical 
exponents and minimize the scatter of the data collapse
plot, the resulting $\gamma/\nu$ is within 5\% of the 2D Ising value.
An example of such an analysis 
(for the honeycomb lattice) is given in \cite{zhang19}.

\begin{figure*}[t!]
\centering
\hskip-0.50in
\includegraphics[height=2.8in,width=7.5in]{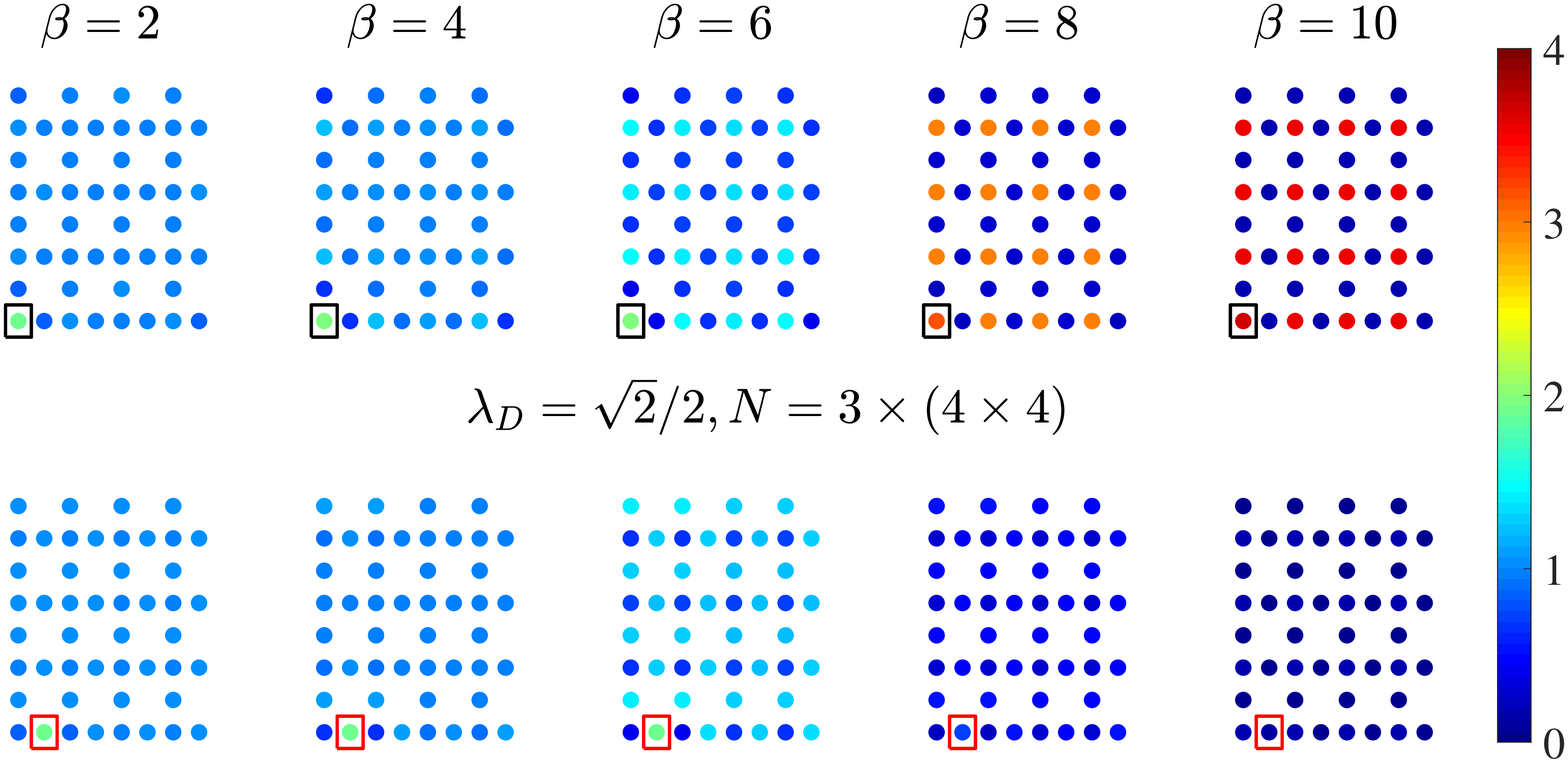}
\caption{
Density-density correlation for a $3\times(4\times4)$ Lieb lattice at
$\omega_0=1, \lambda=2 (\lambda_D=\sqrt{2}/2)$.  The simulation 
was initialized with phonon displacement field appropriate to being in the
$\rho=1/3$ minimum with dominant $A$ sublattice 
(`Copper sites') occupation.
\underbar{First row:}  correlations between each site and 
the Cu site in the bottom left unit cell. 
\underbar{Second row:} 
correlations between each site and the $B/C$ sublattice
(`Oxygen sites') in the  bottom left  unit cell.
}
\label{fig:dd_corr_diff_beta_rand1} 
\end{figure*}

\begin{figure*}[!htb]
\centering
\hskip-0.50in
\includegraphics[height=2.8in,width=7.5in]{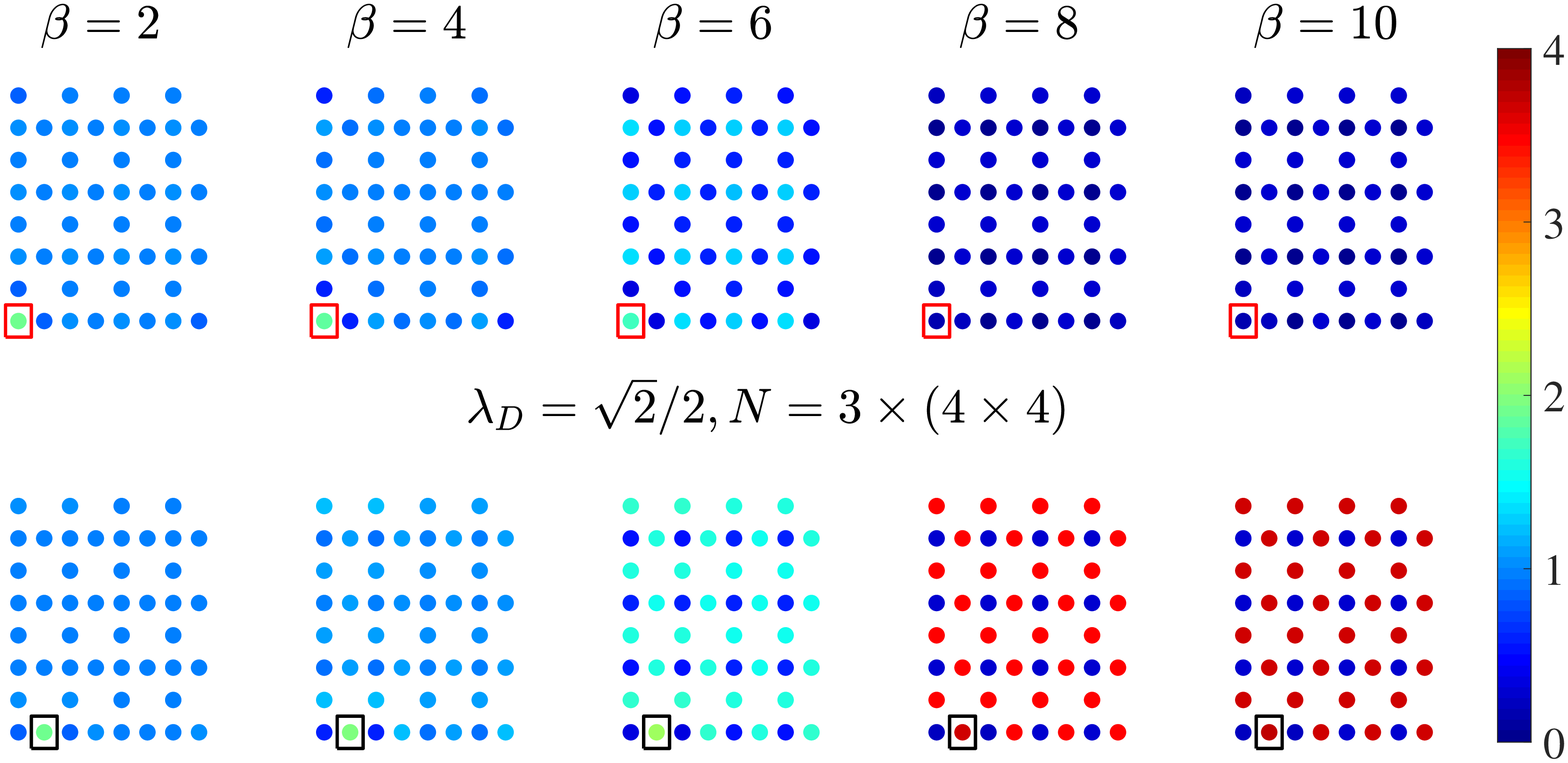}
\caption{
Same as Fig.~\ref{fig:dd_corr_diff_beta_rand1} 
except starting in the $\rho=2/3$ minimum.
}
\label{fig:dd_corr_diff_beta_rand2} 
\end{figure*}

The real space density correlations $c(\mathbf{r})$ 
provide additional insight into the nature of the CDW
order.   Figures
\ref{fig:dd_corr_diff_beta_rand1}  and \ref{fig:dd_corr_diff_beta_rand2} 
give color intensity plots of 
$c(\mathbf{r})$ for different temperatures and initializations
of the phonon displacement $x(\mathbf{i},\tau)$.  
More specifically, if we start the phonon
displacement at $x_0 - \Delta$ (with $\Delta>0$), the fermion density on
that site tends to be high, while a displacement $x_0 + \Delta$ is associated
with a low density.
At high temperatures, the 
correlations are independent of the starting configuration
and $c(\mathbf{r}) = \langle n_{\mathbf{i+r}} n_{\mathbf{i}} \rangle
= \langle n_{\mathbf{i+r}} \rangle 
\langle n_{\mathbf{i}} \rangle \sim 1$.
Short range correlations begin to develop at $\beta\,t \sim 6$ 
and a strong
alternation between $c(\mathbf{r}) \sim 4$, 
where $\mathbf{r}$ connects a pair of doubly occupied sites, 
and $c(\mathbf{r}) \sim 0$, where one of the sites is empty, 
becomes apparent.  
In the case of the initialization in the $\rho=1/3$ 
state (Fig.~\ref{fig:dd_corr_diff_beta_rand1}) with 
only sublattice A sites occupied,
density correlations referenced to an A site (top panel) 
show the alternation,
whereas if referenced to an unoccupied B site 
(bottom panel) all $c(\mathbf{r})$ become small.
Conversely, for initialization in the $\rho=2/3$ state
(Fig.~\ref{fig:dd_corr_diff_beta_rand2}) 
with sublattice B,C sites occupied,
density correlations referenced to a B site (bottom panel) 
show the alternation, whereas if referenced to an 
unoccupied A site (top panel) all $c(\mathbf{r})$ become small.

\begin{figure}[t!]
\includegraphics[height=2.35in,width=3.2in]{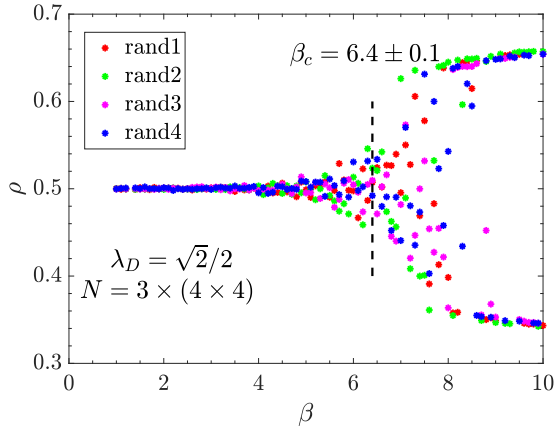}
\caption{
Density per spin $\rho$ as a function of $\beta$ at 
the $\lambda_D=\sqrt{2}/2$. 
Data for four different random seeds are shown. 
A spontaneous symmetry breaking begins to occur at $\beta \sim 5$.
See text for details.
The vertical dashed line is the value of $\beta_c$ 
determined from FSS of $S_{\rm cdw}$.
}
\label{fig:density_beta} 
\end{figure}

Another way to examine the evolution into 
one of two possible ground states,
characterized by distinct densities, is to begin several simulations
with constant density $\rho=1/2$ per spin, and 
examine the final densities achieved.
Figure \ref{fig:density_beta} shows the result for four such
simulations.  At small $\beta$ the lattice remains half-filled,
but as $\beta$ increases the lattice falls into either the
$\rho=1/3$ or the $\rho=2/3$ minimum.  
The tendency for this splitting begins
about $\beta \sim 5$.
For $5 \lesssim \beta \lesssim 9$ the data tend 
to fill the region between
the upper and lower densities.  
This happens because at finite temperatures
and on finite lattices, tunneling between the 
two minima can occur in the course
of a simulation.  Depending on the relative amount of time spent 
at $\rho=1/3$ and $\rho=2/3$, the average density 
can take different values.
For $\beta \gtrsim 9$ very little tunneling occurs, and the data
instead lie on just one of the two bounding lines.  
Note that  the order parameter depends on $\beta$ so that the
increasing width of the $\rho$ curves reflects 
the growth of the CDW order
parameter below $\beta_c$.

It is important to emphasize a subtlety of the physics.  Although the
simulations of Fig.~\ref{fig:density_beta}  
were done at the chemical potential
$\mu = -\lambda^2/\omega_0^2$ which 
should give $\rho=1/2$ per spin by particle-hole symmetry,
the symmetry is broken and there are two low temperature phases with
$\rho=1/3$ and $\rho=2/3$.  This is precisely 
analogous to a simulation 
of a magnetic (e.g.~Ising) model at zero external field.  Although
symmetry demands magnetization $M=0$, 
below $T_c$ there are two phases with $M=\pm M_*$.

\begin{figure}[t!]
\includegraphics[height=2.35in,width=3.2in]{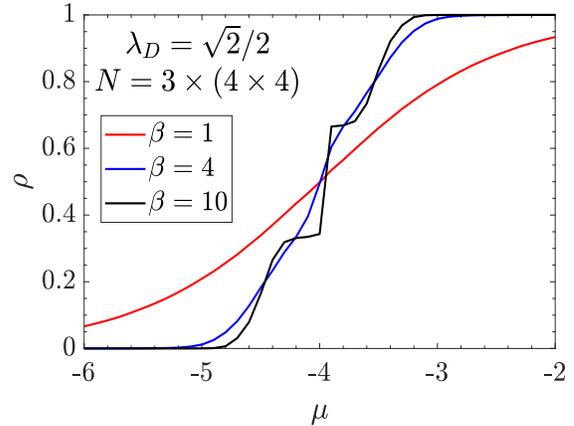}
\caption{
Density per spin $\rho$ vs.~chemical potential $\mu$ for several
different $\beta$ obtained in DQMC simulations. Here $\lambda=2, (\lambda_D=\sqrt{2}/2)$. 
}
\label{fig:density_mu_diffbeta} 
\end{figure}

Plots of the density  $\rho$ versus chemical potential $\mu$
(Fig.~\ref{fig:density_mu_diffbeta})
also reveal the CDW phase.  At high 
temperatures $\rho$ evolves smoothly
between the empty and a fully-packed limits, transiting half-filling
at the particle-hole symmetry point $\mu = -\lambda^2/\omega_0^2$.
At temperatures below the CDW transition, a plateau develops
in which the compressibility $\kappa=d\rho/d\mu$ vanishes.
However, unlike the situation on a bipartite lattice 
in which each sublattice
has equal numbers of particles, the plateau is bifurcated
by an abrupt jump as the system transitions from occupation of the
minority to majority sublattice.

\begin{figure}[t!]
\includegraphics[height=2.35in,width=3.2in]{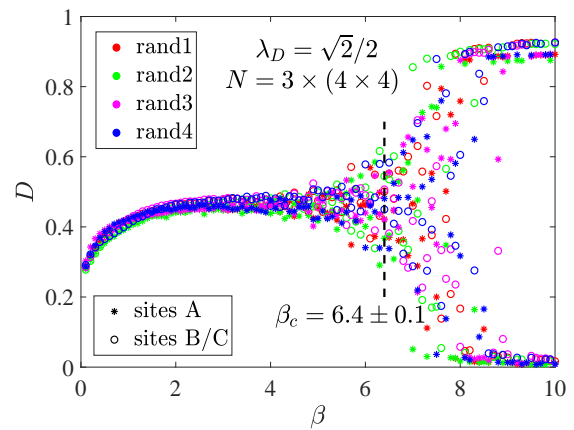}
\caption{ 
Double occupancy ${\cal D}$ vs $\beta$ at $\lambda=2
(\lambda_D=\sqrt{2}/2)$ for a 3*(4*4) lattice and 
$\mu = - \lambda^2/\omega_0^2$.  
Data for four different random seeds are shown. 
The vertical dashed line is the value of $\beta_c$ 
determined from FSS of $S_{\rm cdw}$. At high T (small $\beta$), 
electrons are uncorrelated, and $D \approx \langle n_{{\bf i}\uparrow}\rangle 
\langle n_{{\bf i}\downarrow} \rangle \sim 0.25$ on every site ${\bf i}$. As $T$ decreases, 
pairs begin to form on half the sites, leaving the other half empty, and the double occupancy increases
to $D \sim 0.5$. Finally, as $T$ is further lowered, below $1/\beta_c$, 1/3 and 2/3 filling 
CDW patterns are revealed, with distinct values of ${\cal D}$ on the two sublattices, reflecting
spontaneous symmetry breaking. 
}
\label{fig:double_occupancy} 
\end{figure}

Figure \ref{fig:double_occupancy} is similar to 
Fig.~\ref{fig:density_beta} 
except showing the double occupancy ${\cal D}$.
At low $\beta$ (high $T$),
${\cal D} = \langle n_{\mathbf{i}\uparrow} n_{\mathbf{i}\downarrow} \rangle \sim
\langle n_{\mathbf{i}\uparrow}\rangle \langle n_{\mathbf{i}\downarrow} \rangle 
\sim 1/4$.
As $T$ decreases below the pair binding scale
$U_{\rm eff} = \lambda^2/\omega_0^2 \sim 4$, pairs begin to form on half the sites
(${\cal D} \sim 0.5$).  At larger $\beta$ a CDW pattern emerges
in which ${\cal D}=0$ or ${\cal D}=1$ depending on 
which sublattice is occupied.

\begin{figure}[t!]
\includegraphics[height=2.35in,width=3.2in]{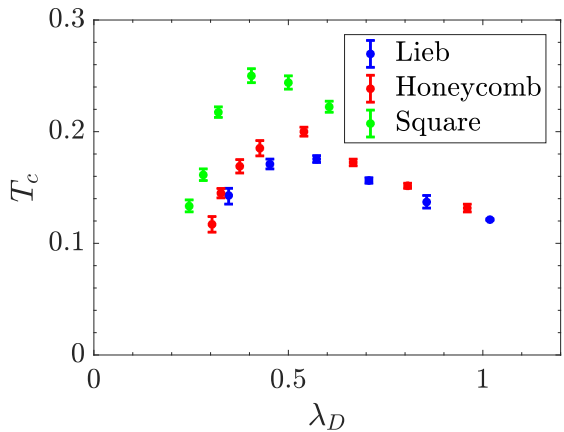}
\caption{
Critical temperatures for the Lieb lattice (this work) and
the honeycomb \cite{zhang19} and square lattices.
}
\label{fig:Tc_lambda_D} 
\end{figure}

Figure \ref{fig:Tc_lambda_D} is the phase diagram of 
the Holstein model
on a Lieb lattice in the plane of temperature-dimensionless 
coupling constant.
We also compare to several other geometries.  
A striking feature of the plot is
that the honeycomb and Lieb lattice values are 
so close.  Naively, one might have 
argued that the delta-function divergence of the 
Lieb lattice flat band density
of states would lead to a large $T_c$, especially 
when compared to the semi-metallic
case of the honeycomb lattice.  However, the explanation 
is clear-  The 
Lieb lattice CDW order really occurs for 
$\rho=1/3$ and $\rho=2/3$, where 
it has Dirac cones much like the honeycomb lattice.  
Thus the only difference is
that the honeycomb lattice coordination number $z=3$, 
whereas for the Lieb
geometry the average coordination number is 
slightly smaller
$\bar z = 2/3 (2) + 1/3(4) = 8/3$.  
Obtaining the weak coupling behavior of $T_c$ is a nontrivial analytic 
calculation.  It has been done for the 2D square lattice, yielding good agreement
with DQMC simulations similar to those reported here\cite{freericks94}.
\color{black}

%%%%%%%%%%%%%%%%%%%%%%%%%%%%%%%%%%%%%%%%%%%%%%%%%%%%%%%%%%%%%%%%%%
\section{Conclusions}  \label{sec:Conclusions}
%%%%%%%%%%%%%%%%%%%%%%%%%%%%%%%%%%%%%%%%%%%%%%%%%%%%%%%%%%%%%%%%%%

We have studied the charge density wave transition for 
the Holstein model
on a Lieb lattice.  Our interest was in establishing results for the 
effect of compact localized states (flat bands) 
on ordered phases driven by
the electron-phonon interaction, in analogy with the 
body of work which 
exists for electron-electron interactions 
(primarily the Hubbard model).

The behavior of the occupation, double occupation, spectral function,
and charge structure factor have been obtained quantitatively, and
used to infer a phase diagram of critical temperature versus
coupling constant.

We emphasize as well that our results for
electron-phonon interactions on a Lieb lattice
differ from those for electron-electron
interactions\cite{iglovikov14} in a fundamental way.  
The degeneracy of the superconducting and CDW orders at half-filling
in the half-filled attractive Hubbard model implies the absence of long range 
order except in the ground state (Mermin-Wagner).
This symmetry is broken  in the Holstein 
model.  As a consequence there is a finite CDW $T_c$ 
even on two dimensional geometries.
This is already well-known for the square and honeycomb lattices.
\color{black}

%%%%%%%%%%%%%%%%%%%%%%%%%%%%%%%%%%%%%%%%%%%%%%%%%%%%%%%%%%%%%%%%%%
%\section*{ACKNOWLEDGMENTS}
%%%%%%%%%%%%%%%%%%%%%%%%%%%%%%%%%%%%%%%%%%%%%%%%%%%%%%%%%%%%%%%%%%
%% \begin{acknowledgments}
\vskip0.10in
\noindent
\underbar{\bf Acknowledgements:}
The work of C.F.~and R.S.~was supported by the 
grant DE‐SC0014671 funded by
the U.S. Department of Energy, Office of Science.

\bibliography{Feng_Lieb}

\end{document}